\documentclass{ap-jnmp}

\copyrightauthor{}

\title{A reciprocal transformation for a  3-component Camassa-Holm type system}

\author{Nianhua Li\footnote{%
Corresponding author.}}

\address{School of Mathematics,
Huaqiao University\\
Quanzhou, Fujian 362021, People's Republic of China\\
\email{linianh@hqu.edu.cn}}

\begin{document}

\maketitle
\thispagestyle{empty}

\vphantom{\vbox{%
\begin{history}
\received{(Day Month Year)}
\revised{(Day Month Year)}
\accepted{(Day Month Year)}
\end{history}
}}

\begin{abstract}
A reciprocal transformation for a 3-component Camassa-Holm type system is constructed to connect it with the first negative flow of a generalized MKdV hierarchy, and a bi-Hamiltonian structure for the transformed system is also considered.
\end{abstract}

\ccode{2000 Mathematics Subject Classification: 37K10, 35Q51, 35Q58}

\section{Introduction}
Since the celebrated Camassa-Holm (CH) equation was derived as a model for unidirectional motion of dispersive shallow-water by Camassa and Holm in 1993  \cite{Holm,Hyman},
integrable PDEs admitting peakons (CH type equations) have attracted much attention in recent years. Two of the most famous
equations are the CH equation and the Degasperis-Procesi (DP) equation, which are contained in the
following one parameter family of PDEs \cite{gas}:
\begin{equation}\label{ch}
m_t+um_x+bu_xm=0,\quad m=u-u_{xx},
\end{equation}where $b$ is an arbitrary constant. When $b = 2$, the equation is the CH, which is completely integrable
with a Lax pair and a bi-Hamiltonian structure given \cite{Holm,Hyman}.
The CH equation is solvable by inverse scattering transform \cite{Con1} and the associated algebro-geometric solutions are also discussed \cite{Gez}. And explicit formulas for the multi-peakon solutions of the CH equation are studied by the inverse spectral methods\cite{beals1,beals2}. Moreover, it is connected with the negative KdV equation by a reciprocal transformation \cite{fuch,lene,hon2}. When $b=3$, the system (\ref{ch}) is just the DP equation,  which is discovered by the method of asymptotic integrability to isolate integrable third-order equations \cite{Deg}. Indeed it may also be obtained by the theory of shallow water \cite{dulin}. The DP equation is connected to a negative flow in the Kaup-Kupershmidt hierarchy via a reciprocal transformation, and a Lax pair of the equation is constructed by the relation. The bi-Hamiltonian structure and the integrable finite-dimensional peakon dynamics for the DP equation are also studied in \cite{gas}. Furthermore, the research of explicit formulas for the multi-peakon solutions of the DP equation can be found in \cite{lun1,lun2}.

In the past two decades, many other CH type equations are proposed and studied (see e.g. \cite{wang,popo} and references). For example, the Novikov equation \cite{Novikov}
\begin{equation*}
m_t+u^2m_x+3uu_xm=0,\quad m=u-u_{xx}
\end{equation*} is discovered by the method of symmetry classification. Bi-hamiltonian structure and a Lax pair for the Novikov equation are given and it is shown to connect with a negative flow in the Sawada-Kotera hierarchy \cite{Hone}. And explicit formulas for multi-peakon solutions of it are obtained in \cite{ahon}.
In addition, the Geng-Xue equation
\begin{eqnarray*}
&&m_t+3u_xvm+uvm_x=0,\\
&&n_t+3uv_xn+uvn_x=0,\\
&& m=u-u_{xx},\quad  n=v-v_{xx}
\end{eqnarray*}is proposed as a generalization of the Novikov equation \cite{Geng}. It is completely integrable with a Lax pair and associated bi-Hamiltonian structure as well as infinitely many conserved quantities\cite{Geng,ll}. Furthermore, a reciprocal transformation and explicit formulas for the multi-peakon of it are discussed in \cite{li3,ldk}.

The subject of this paper is a three-component CH type system proposed by Geng and Xue \cite{Geng2}, i.e.
\begin{eqnarray}\label{gengxue3}
&&u_{t}=-v p_{x}+u_{x}q+\frac{3}{2}u q_{x}-\frac{3}{2} u(p_{x}r_{x}-pr),\nonumber\\
&&v_{t}=2vq_{x}+v_{x}q ,\label{h1}\\
&&w_{t}=v r_{x}+w_{x} q+\frac{3}{2}w q_{x}+\frac{3}{2} w(p_{x}r_{x}-pr),\nonumber
\end{eqnarray}
where
\begin{eqnarray*}
&&u=p-p_{xx},\nonumber\\
&&v=\frac{1}{2}(q_{xx}-4q+p_{xx}r_{x}-r_{xx}p_{x}+3p_{x}r-3p r_{x}),\\
&&w=r_{xx}-r,\nonumber
\end{eqnarray*}which admits the following spectral problem
 \begin{equation}\label{thr2}
\psi_x=\left(
         \begin{array}{ccc}
           0 & 1 & 0 \\
           1+\lambda v &0 & u \\
           \lambda w & 0 & 0 \\
         \end{array}
       \right)\psi.
\end{equation} Bi-Hamiltonian structure as well as infinite many conserved quantities and the dynamical system for the $N$-peakon solutions of this system are discussed in\cite{Geng2,Li}. It is interesting that the spectral problem (\ref{thr2}) for this 3-component model may be reduced to which of the CH equation and the Geng-Xue equation, and bi-Hamiltonian structures of the two equations can also be obtained by reductions for bi-Hamiltonian structure of the system (\ref{gengxue3}).
We are intrigued about the reciprocal transformation of this system, because it turns out that the spectral problem (\ref{thr2}) for the 3-component system (\ref{gengxue3}) is gauge link to a spectral problem of a second three-component CH type system constructed by us in \cite{popo}, however, the reciprocal transformation (more precisely, Liouville transformation) on our 3-component CH type system can't be applied to the system (\ref{gengxue3}) \cite{Li2}.

The aim of this paper is to construct a reciprocal transformation for the 3-component system (\ref{gengxue3}). The results show that the transformed system is linked to a negative flow in a 3-component extension to the MKdV hierarchy.   
\section{Reduction of the spectral problem}
It is easy to find that the spectral problem (\ref{thr2}) is reduced to which of the CH equation in the condition $u=w=0$, and reduced to which of the Geng-Xue equation as $v=0$. Meanwhile, we note that bi-Hamiltonian structure of the 3-component system is written as
\begin{equation}\label{biham1}
\left(
  \begin{array}{c}
    u \\
    w \\
    v \\
  \end{array}
\right)_t={\cal J}_1\left(
                    \begin{array}{c}
                      \frac{\delta H_1}{\delta u} \\
                      \frac{\delta H_1}{\delta w} \\
                      \frac{\delta H_1}{\delta v} \\
                    \end{array}
                  \right)={\cal J}_2\left(
                    \begin{array}{c}
                      \frac{\delta H_0}{\delta u} \\
                      \frac{\delta H_0}{\delta w} \\
                      \frac{\delta H_0}{\delta v} \\
                    \end{array}
                  \right)
\end{equation}where
\begin{equation*}
{\cal J}_1=\left(
           \begin{array}{ccc}
             0 & \partial^2-1 & 0 \\
             1-\partial^2 & 0 & 0 \\
             0 & 0 &  -\partial v-v\partial \\
           \end{array}
         \right),\quad {\cal J}_2={\cal M}-{\cal N}(\partial^3-4\partial)^{-1}{\cal N}^{*}
\end{equation*}with
\begin{equation*}
 {\cal M}=\left(
                             \begin{array}{ccc}
                               \frac{3}{2}u\partial^{-1}u & -v-\frac{3}{2}u\partial^{-1}w & 0 \\
                               v-\frac{3}{2}w\partial^{-1} u & \frac{3}{2}w\partial^{-1}w & 0\\
                               0 & 0 & 0 \\
                             \end{array}
                           \right),\quad {\cal N}=\left(
           \begin{array}{c}
             \frac{3}{2}u\partial+u_x \\
             \frac{3}{2}w\partial+w_x \\
             v\partial+\partial v\\
           \end{array}
         \right)
\end{equation*}and
\begin{eqnarray*}
  H_1 &=& \frac{1}{4}\int [4q^2-qq_{xx}-p_x^2r_x^2+6pp_xrr_x+3p^2r^2]dx, \\
  H_0 &=& \int [v+ur_x] dx.
\end{eqnarray*}
Direct calculations show that the compatible Hamiltonian operators for the CH and the Geng-Xue may be obtained by the reduction of the Hamiltonian pair in (\ref{biham1}) in the case of $u=w=0$ and $v=0$ respectively. Besides it is easy to find that the Hamiltonian pair and a spectral problem for the Novikov hierarchy can also be obtained when $u=w,v=0$.

\section{A reciprocal transformation}
It was shown in \cite{Geng2} that a Lax pair of the 3-component system (\ref{gengxue3}) is given by the spectral problem (\ref{thr2}) and associated  auxiliary problem
\begin{equation}\label{tpart}
\psi_t=\left(
         \begin{array}{ccc}
           -\frac{1}{2}(q_x+p_xr_x-pr) & \frac{1}{\lambda}+q & \frac{p_x}{\lambda} \\
           \frac{1}{\lambda}-pr_x+p_xr-q+\lambda qv & \frac{1}{2}(q_x-p_xr_x+pr) & \frac{p}{\lambda}+qu \\
           -r+\lambda qw & r_x & p_xr_x-pr \\
         \end{array}
       \right)\psi.
\end{equation}Based on this Lax pair, infinitely many conservation laws for the 3-component system (\ref{gengxue3}) have been constructed\cite{Geng2}. Specially, one of them is
\begin{equation*}
(v^{\frac{1}{2}})_t=(v^{\frac{1}{2}}q)_x,
\end{equation*}which defines a reciprocal transformation
\begin{equation}\label{rt1}
dy=v^{\frac{1}{2}}dx+v^{\frac{1}{2}}qdt,\quad d\tau=dt.
\end{equation}

In what follows, the 3-component CH system (\ref{gengxue3}) with the Lax pair (\ref{thr2}-\ref{tpart}) under the reciprocal transformation (\ref{rt1}) will be shown. Writing the column vector $\psi$ in components as $\psi=(\psi_1,\psi_2,\psi_3)^{T}$ and eliminating $\psi_2$ in the spectral problem (\ref{thr2}) , we find
\begin{equation*}
\psi_{1xx}=(1+\lambda v)\psi_1+u\psi_3,\quad \quad \quad \psi_{3x}=\lambda w \psi_1.
\end{equation*}Under the reciprocal transformation (\ref{rt1}) and after the gauge transformation $\psi_1=v^{-\frac{1}{4}}\phi_1$, the above spectral problem is transformed to
\begin{equation*}
 \phi_{1yy}+(\frac{3v_y^2}{16v^2}-\frac{v_{yy}}{4v}-\frac{1}{v})\phi_1=\lambda\phi_1+uv^{-\frac{3}{4}}\psi_3,\quad \psi_{3y}=\lambda wv^{-\frac{3}{4}}\phi_1.
\end{equation*}Using the method of factorizing the Lax operator \cite{fordy}, we arrive at the following spectral problem:
\begin{equation}\label{tranlax1}
\phi_y=\left(
                  \begin{array}{ccc}
                    Q_1 & \mu & 0 \\
                    \mu & -Q_1 & Q_2\\
                    \mu Q_3 & 0 & 0 \\
                  \end{array}
                \right)\phi,\quad \quad \quad  \phi=\left(
                                         \begin{array}{c}
                                           \phi_1 \\
                                           \phi_2 \\
                                           \phi_3 \\
                                         \end{array}
                                       \right),
\end{equation}where $\phi_2=\frac{1}{\mu}(\partial_y-Q_1)\phi_1,\phi_3=\frac{1}{\mu}\psi_3$ with $\mu=\lambda^{\frac{1}{2}}$ and
\begin{equation*}
Q_1=\frac{v_y}{4v}+v^{-\frac{1}{2}},\quad Q_2=uv^{-\frac{3}{4}},\quad Q_3=wv^{-\frac{3}{4}}.
\end{equation*}
It is worth to note that many choices for the transformed spectral problems are tried here but failed since the variable $q$ is lost in the time part of the Lax pairs corresponding to the transformed spectral problems. The transformed spectral problem (\ref{tranlax1}) can be reduced to which of the MKdV hierarchy as $Q_2=Q_3=0$ and we call the associated hierarchy a generalized MKdV hierarchy.

Furthermore, taking similar procedure as before and through  tedious calculations, we get the auxiliary problem corresponding to the spectral problem (\ref{tranlax1}), that is
\begin{equation}\label{tranlax2}
\phi_{\tau}=\left(
                        \begin{array}{ccc}
                          \frac{1}{\mu^2}-\frac{1}{2}f & \frac{1}{\mu}g & \frac{1}{\mu}((a_y-aQ_1)g+a) \\
                          \frac{1}{\mu}(c-g^{-1}\partial^{-1}(aQ_3+bQ_2))& -\frac{1}{\mu^2}-\frac{1}{2}f & \frac{1}{\mu^2}(aQ_1-a_y) \\
                          \frac{1}{\mu}(b_y-bQ_1) & (b_y-bQ_1)g+b & f \\
                        \end{array}
                      \right)\phi,
\end{equation}where
\begin{equation*}
a=pv^{\frac{1}{4}},\quad \quad b=rv^{\frac{1}{4}},\quad \quad c=q_y-2qv^{-\frac{1}{2}}
\end{equation*}and
\begin{eqnarray*}
&&f=\partial_y^{-1}[(Q_3(a_y-aQ_1)g-Q_2(b_y-bQ_1)g+aQ_3-bQ_2],\\
&&g=-(\partial_y-2Q_1)^{-1}2.
\end{eqnarray*}
Then the compatible condition of the Lax pair (\ref{tranlax1}-\ref{tranlax2}) yields just the transformed system
\begin{equation}\label{treqq}
\left(
  \begin{array}{c}
    Q_{1} \\
    Q_{2} \\
    Q_{3} \\
  \end{array}
\right)_{\tau}=\left(
                 \begin{array}{c}
                   \frac{1}{2}(cg)_y \\
                   -\frac{3}{2}Q_2f-(a_y-aQ_1)g-a \\
                   \frac{3}{2}Q_3f+(b_y-bQ_1)g+b \\
                 \end{array}
               \right),\quad \quad \quad \left(
                               \begin{array}{c}
                                 F_1 \\
                                 F_2\\
                                 F_3 \\
                               \end{array}
                             \right)=0.
\end{equation}where
\begin{eqnarray*}
&&F_1=a_{yy}-a(Q_{1y}+Q_1^2)+Q_2, \nonumber\\
&&F_2=Q_3-b_{yy}+b(Q_{1y}+Q_1^2),\\
&&F_3=Q_2(b_y-bQ_1)+Q_3(a_y-aQ_1)+(g^{-1}+2g^{-2}\partial_y^{-1})(aQ_3+bQ_{2})-c_y-2Q_1c+2.\nonumber
\end{eqnarray*}
Consequently, the 3-component CH type system (\ref{gengxue3}) and the Lax pair (\ref{thr2}-\ref{tpart}) are transformed to (\ref{treqq}) and (\ref{tranlax1}-\ref{tranlax2}) respectively.

\section{Hamiltonian pair and conserved quantities of the generalized MKdV hierarchy}
In this section, we will discuss the bi-Hamiltonian structure of the generalized MKdV hierarchy. One may note that in \cite{gengli} a generalized KdV hierarchy  is proposed associated with a 3 $\times$ 3 matrix spectral problem
\begin{equation}\label{ipsp}
\varphi_x=\left(
            \begin{array}{ccc}
              0 & 1 & 0 \\
              \lambda+v_1 & 0 & u_1 \\
              \lambda w_1 & 0 & 0 \\
            \end{array}
          \right)\varphi,
\end{equation}which may be transformed to a spectral problem of Yajima-Oikawa hierarchy (see e.g. \cite{Liu, Yaji}) or a sepctral problem associated with a energy-dependent Lax operator (example 4.1 in \cite{antono}) by two gauge transformations respectively. By writing the column vector $\varphi$ in components and comparing it with the similar result of (\ref{tranlax1}), we find the relation of the three potentials in the two system is related as
\begin{equation}\label{rela1}
u_1=Q_2,\quad v_1=Q_{1y}+Q_1^2,\quad w_1=Q_3.
\end{equation}
Bi-Hamiltonian structure of the generalized KdV hierarchy associated with the spectral problem (\ref{ipsp}) is given without a proof, since the original expression of it is tedious, and we rewrite the Hamiltonian pair of it as
\begin{eqnarray*}
{\cal E}_1 &=& \left(
                    \begin{array}{ccc}
                      -\frac{3}{2}u_1\partial_y^{-1}u_1 & 0 & \frac{3}{2}u_1\partial_y^{-1}w_1+1 \\
                     0 & -\frac{1}{2}\partial_y^3+\partial v_1+v_1\partial & 0 \\
                     \frac{3}{2}w_1\partial_y^{-1}u_1-1 & 0 & -\frac{3}{2}w_1\partial_y^{-1}w_1 \\
                    \end{array}
                  \right),\\
{\cal E}_2 &=& \left(
                   \begin{array}{ccc}
                     0 & 0 & \partial_y^2-v_1 \\
                     0 & 0 & 0 \\
                     v_1- \partial_y^2 & 0 & 0 \\
                   \end{array}
                 \right)+\frac{1}{2}\left(
                                      \begin{array}{c}
                                        \frac{1}{2}u_1\partial_y+\partial_y u_1 \\
                                        -\frac{1}{2}\partial_y^3+\partial_y v_1+v_1\partial_y \\
                                        \frac{1}{2}w_1\partial+\partial_y w \\
                                      \end{array}
                                    \right)\partial_y^{-1}\left(
                                      \begin{array}{c}
                                        \frac{1}{2}u_1\partial_y+\partial_y u_1 \\
                                        -\frac{1}{2}\partial_y^3+\partial_y v_1+v_1\partial_y \\
                                        \frac{1}{2}w_1\partial+\partial_y w \\
                                      \end{array}
                                    \right)^{*}.
\end{eqnarray*}
Through the relation (\ref{rela1}),  we may get the compatible Hamiltonian operators and associated Hamiltonian functions for the generalized MKdV hierarchy associated with the spectral problem (\ref{tranlax1}). Since the method of trace identity hasn't been taken, we only give the Hamiltonian pair here.
{\theorem}The generalized MKdV hierarchy associated with the spectral problem (\ref{tranlax1}) has a pair of Hamiltonian operators
\begin{eqnarray*}
  {\cal K}_1 &=& \left(
                         \begin{array}{ccc}
                           -\frac{1}{2}\partial_y & 0 & 0 \\
                           0 & \frac{3}{2}Q_2\partial_y^{-1}Q_2 & -1-\frac{3}{2}Q_2\partial_y^{-1}Q_3 \\
                           0 & 1-\frac{3}{2}Q_3\partial_y^{-1}Q_2 & \frac{3}{2}Q_3\partial_y^{-1}Q_3 \\
                         \end{array}
                       \right), \\
  {\cal K}_2 &=& \left(
                         \begin{array}{ccc}
                           0 & 0 & 0\\
                           0 & 0 & (\partial_y+Q_1)(\partial_y-Q_1) \\
                           0 & -(\partial_y+Q_1)(\partial_y-Q_1) & 0 \\
                         \end{array}
                       \right)+\frac{1}{2}{\cal F}\partial_y^{-1}{\cal F}^{*},
\end{eqnarray*}where
\begin{equation*}
{\cal F}=(-\frac{1}{2}\partial_y(\partial_y-2Q_1),\frac{1}{2}Q_2\partial_y+\partial_y Q_2,\frac{1}{2}Q_3\partial_y+\partial_y Q_3)^{T}.
\end{equation*}
Proof: In order to prove the theorem, we need to show the two operators $ {\cal K}_1$ and ${\cal K}_2$ are Hamiltonian operators and they constitute a compatible pair. It is sufficient to show $a{\cal K}_1+b{\cal K}_2$ is a Hamiltonian operator for arbitrary constants $a$ and $b$. Setting ${\cal L}=a{\cal K}_1+b{\cal K}_2$,  one can easily show that ${\cal L}$ is skew-symmetric and we only need to prove
\begin{equation*}
{\rm pr\ v}_{{\cal L}\theta}\Theta_{{\cal L}}=0.
\end{equation*}
Let $e=\partial_y^{-1}(Q_2\theta_2-Q_3\theta_3)$ and
\begin{equation*}
h=\partial_y^{-1}(\frac{1}{2}\theta_{1yy}+Q_1\theta_{1y}+\frac{3}{2}Q_2\theta_{2x}+\frac{1}{2}Q_{2y}\theta_2+\frac{3}{2}Q_3\theta_{3y}+\frac{1}{2}Q_{3y}\theta_3),
\end{equation*}then
\begin{equation}\label{prv}
{\cal L}\theta=\left(
                                            \begin{array}{c}
                                              -\frac{1}{2}a\theta_{1y}+\frac{1}{4}bh_{yy}-\frac{1}{2}b(Q_1h)_y \\
                                              \frac{3}{2}aQ_2e-a\theta_3+b\theta_{3yy}-b(Q_{1y}+Q_1^2)\theta_3-\frac{3}{4}bQ_2h_y-\frac{1}{2}bQ_{2y}h \\
                                               -\frac{3}{2}aQ_3e+a\theta_2-b\theta_{2yy}+b(Q_{1y}+Q_1^2)\theta_2-\frac{3}{4}bQ_3h_y-\frac{1}{2}bQ_{3y}h \\
                                            \end{array}
                                          \right)=\left(
                                                    \begin{array}{c}
                                                      L_1 \\
                                                      L_2 \\
                                                      L_3 \\
                                                    \end{array}
                                                  \right),
\end{equation}and
\begin{eqnarray*}
&& \Theta_{{\cal L}}=\frac{1}{2}\int \theta\wedge {\cal L}\theta dy\\
&& \hspace{0.7cm}=\int [\frac{3}{4}a(Q_2\theta_2-Q_3\theta_3)\wedge e-\frac{1}{4}a\theta_1\wedge \theta_{1y}-a\theta_2\wedge\theta_3-b(Q_{1y}+Q_1^2)\theta_2\wedge \theta_3+b\theta_2\wedge \theta_{3yy}\\
&& \hspace{1.4cm}+\frac{1}{4}b(\frac{1}{2}\theta_{1yy}+Q_1\theta_{1y}+\frac{3}{2}Q_2\theta_{2y}+\frac{1}{2}Q_{2y}\theta_2+\frac{3}{2}Q_3\theta_{3y}+\frac{1}{2}Q_{3y}\theta_3)\wedge h]dy,
\end{eqnarray*}therefore
\begin{eqnarray*}
&&{\rm pr\ v}_{{\cal L}\theta}\Theta_{{\cal L}}=\int [\frac{3}{2}a(L_2\wedge \theta_2-L_3\wedge \theta_3)\wedge e-b(L_{1y}+2Q_1L_1)\wedge \theta_2\wedge\theta_3\\
&&\hspace{2.2cm}+\frac{1}{2}b(L_1\wedge \theta_{1y}+\frac{3}{2}L_2\wedge \theta_{2y}+\frac{1}{2}L_{2y}\wedge \theta_{2}+\frac{3}{2}L_3\wedge \theta_{3y}+\frac{1}{2}L_{3y}\wedge \theta_{3})\wedge h ]dy\\
&&\hspace{1.8cm}=\int [\frac{3}{2}aL_2\wedge \theta_2\wedge e-\frac{3}{2}aL_3\wedge \theta_3\wedge e+bL_1\wedge (\theta_{2y}\wedge \theta_3+\theta_2\wedge \theta_{3y})-2bQ_1L_1\wedge \theta_2\wedge \theta_3\\
&&\hspace{2.2cm}+\frac{1}{2}bL_1\wedge \theta_{1y}\wedge h+\frac{3}{4}bL_2\wedge \theta_{2y}\wedge h-\frac{1}{4}bL_2\wedge (\theta_{2y}\wedge h+\theta_2\wedge h_y)+\frac{3}{4}bL_3\wedge \theta_{3y}\wedge h\\
&&\hspace{2.2cm}-\frac{1}{4}bL_3(\theta_{3y}\wedge h+\theta_3\wedge h_y)]dy\\
&&\hspace{1.8cm}=\int [bL_1\wedge (\theta_{2y}\wedge \theta_3+\theta_2\wedge \theta_{3y}-2Q_1\theta_2\wedge \theta_3+\frac{1}{2}\theta_{1y}\wedge h)+L_2\wedge (\frac{3}{2}a\theta_2\wedge e+\frac{1}{2}b\theta_{2y}\wedge h\\
&&\hspace{2.2cm}-\frac{1}{4}b\theta_{2}\wedge h_y)+L_3\wedge (\frac{1}{2}b\theta_{3y}\wedge h-\frac{3}{2}a\theta_3\wedge e-\frac{1}{4}b\theta_{3}\wedge h_y)]dy.
\end{eqnarray*}
Substituting (\ref{prv}) into it and using the basic properties of wedge product, we get
\begin{eqnarray*}
&&{\rm pr\ v}_{{\cal L}\theta}\Theta_{{\cal L}}=\int [-\frac{1}{4}b^2h_y\wedge (\frac{1}{2}\theta_{1yy}+Q_1\theta_{1y}+\frac{3}{2}Q_2\theta_{2x}+\frac{1}{2}Q_{2y}\theta_2+\frac{3}{2}Q_3\theta_{3y}+\frac{1}{2}Q_{3y}\theta_3)\wedge h\\
&&\hspace{2.4cm}+ab\theta_3\wedge \theta_2\wedge h_y+\frac{3}{4}ab(Q_2\theta_2-Q_3\theta_3)\wedge e_y\wedge h+ab(\frac{1}{2}\theta_{1yy}+Q_1\theta_{1y})\wedge \theta_2\wedge \theta_3\\
&&\hspace{2.4cm}-\frac{3}{2}ab(\theta_{2y}\wedge \theta_3+\theta_{3y}\wedge \theta_2)\wedge e_y]dy\\
&&\hspace{1.8cm}=\int [-\frac{1}{4}b^2h_y\wedge h_y\wedge h+ab(\frac{1}{2}\theta_{1yy}+Q_1\theta_{1y}+\frac{3}{2}Q_2\theta_{2y}+\frac{3}{2}Q_3\theta_{3y})\wedge \theta_2\wedge \theta_3\\
&&\hspace{2.4cm}+\frac{3}{4}abe_y\wedge e_y\wedge h-abh_y\wedge \theta_2\wedge \theta_3]dy\\
&&\hspace{1.8cm}=\int [ab(\frac{1}{2}\theta_{1yy}+Q_1\theta_{1y}+\frac{3}{2}Q_2\theta_{2y}+\frac{3}{2}Q_3\theta_{3y}-h_y)\wedge \theta_2\wedge \theta_3]dy\\
&&\hspace{1.8cm}=0.
\end{eqnarray*}Therefore the theorem is proven.

\section{Relation between the transformed system (\ref{treqq})  and the generalized MKdV hierarchy}
We start with the first negative flow in the generalized MKdV hierarchy associated with the spectral problem (\ref{tranlax1}),
\begin{equation}
{\cal K}_2{\cal K}^{-1}_1\left(
                           \begin{array}{c}
                             Q_1 \\
                             Q_2\\
                             Q_3 \\
                           \end{array}
                         \right)_{\tau}=0,
\end{equation}which may be rewritten as
\begin{equation}\label{negflow}
\left(
  \begin{array}{c}
    Q_1 \\
    Q_2 \\
    Q_3 \\
  \end{array}
\right)_{\tau}={\cal K}_1\left(
                           \begin{array}{c}
                             A \\
                             B \\
                             C \\
                           \end{array}
                         \right),\quad \quad \quad {\cal K}_2\left(
                           \begin{array}{c}
                             A \\
                             B \\
                             C \\
                           \end{array}
                         \right)=0.
\end{equation}
In order to find the relation between the transformed system (\ref{treqq}) and the negative flow (\ref{negflow}), we may set
\begin{equation*}
A=-cg,\quad \quad B=(b_y-bQ_1)g+b,\quad \quad C=(a_y-aQ_1)g+a.
\end{equation*} Using $g_y=2Q_1g-2$, we may get the first negative flow for the generalized MKdV hierarchy in the form
\begin{equation*}
\left(
  \begin{array}{c}
    Q_1 \\
    Q_2 \\
    Q_3 \\
  \end{array}
\right)_{\tau}=\left(
                 \begin{array}{c}
                   \frac{1}{2}(cg)_y \\
                   -\frac{3}{2}Q_2f-(a_y-aQ_1)g-a \\
                   \frac{3}{2}Q_3f+(b_y-bQ_1)g+b \\
                 \end{array}
               \right)
,\quad \quad \quad \left(
                           \begin{array}{c}
                            G_1\\
                            G_2 \\
                            G_3\\
                           \end{array}
                         \right)=0,
\end{equation*}
where
\begin{equation*}
\left(
                           \begin{array}{c}
                            G_1\\
                            G_2 \\
                            G_3\\
                           \end{array}
                         \right)=\left(
  \begin{array}{c}
    \frac{1}{4}\partial_y(\partial_y-2Q_1)\partial_y^{-1}[(\partial_yg-\frac{1}{2}g\partial_y)F_3+Q_3gF_1-Q_2gF_2] \\
    \frac{3}{2}g_yF_1+gF_{1y}-\frac{1}{2}(\frac{3}{2}Q_2\partial_y+Q_{2y})\partial_y^{-1}[(\partial_yg-\frac{1}{2}g\partial_y)F_3+Q_3gF_1-Q_2gF_2] \\
    \frac{3}{2}g_yF_2+gF_{2y}-\frac{1}{2}(\frac{3}{2}Q_3\partial_y+Q_{3y})\partial_y^{-1}[(\partial_yg-\frac{1}{2}g\partial_y)F_3+Q_3gF_1-Q_2gF_2]\\
  \end{array}
\right).
\end{equation*}
Then one can find that the transformed system (\ref{treqq}) is a reduction of  the first negative flow (\ref{treqq}) in the generalized MKdV hierarchy.

\bigskip
\noindent
{\bf Acknowledgments}

This work is supported by the National Natural Science Foundation of China (Grant Nos. 11401572, 11401230 and 11505064) and the Initial Founding of Scientific Research for the introduction of talents of Huaqiao University (Project No. 14BS314).

\end{document}